\documentclass[12pt]{iopart}
\usepackage{epsf}
\begin{document}

\title{Anisotropy in the helicity modulus of a quantum 3D XY-model: application to YBCO} 

\author{Bo\v zidar Mitrovi\' c
\footnote[3]{To
whom correspondence should be addressed (mitrovic@newton.physics.brocku.ca)},
Kirill Samokhin, and Shyamal K. Bose }

\address{Physics Department, Brock University, St. Catharines, Ontario,
Canada L2S~3A1}

\begin{abstract}
We present a variational study of the helicity moduli of an 
anisotropic quantum three-dimensional (3D) XY-model of 
YBCO in superconducting state. It is found that both the 
$ab$-plane and the $c$-axis helicity moduli, which are 
proportional to the inverse square of the corresponding 
magnetic field penetration depth, vary with temperature $T$ 
as $T^{4}$ in the zero temperature limit. Moreover, the 
$c$-axis helicity modulus drops with temperature much 
faster than the $ab$-plane helicity modulus because of 
the weaker Josephson couplings along the $c$-axis compared 
to those along the $ab$-plane.
These findings are in disagreement with the 
experiments on high quality samples of YBCO.  
\end{abstract}



\maketitle

\section{Introduction}
In a recent paper [1], following the suggestions of Roddick and Stroud [2] and of 
Emery and Kivelson [3-6], we have examined the possibility that a classical XY-model 
might explain {\em both} the $ab$-plane and the $c$-axis electromagnetic response of 
optimaly doped single crystal YBCO [7]. A Monte Carlo study was applied to a bilayer model for 
YBCO in which all the microscopic degrees of freedom are assumed to be integrated out except for the 
phase of the superconducting order parameter and all of the phase dynamics results from 
Josephson couplings within the layers and between the layers grouped in a stack of bilayers.
We calculated the temperature ($T$) dependence of the helicity moduli which in this model 
correspond to the inverse square of the magnetic field penetration depths $\lambda_{ab}(T)$, 
$\lambda_{c}(T)$, i.e.~to the superfluid densities $n_{s,ab}(T)$, $n_{s,c}(T)$. 
Our results were in sharp contrast with experiment [7]:  
while we found that both $ab$-plane and the $c$-axis superfluid densities decrease linearly 
with $T$ at low temperatures, experimentally only the $ab$-plane superfluid density 
drops linearly with temperature at low $T$. The observed $c$-axis 
penetration depth $\lambda_{c}(T)$ never has the linear temperature dependence found in $ab$-plane.

Here we extend the model of Ref.~[1] to include the quantum effects associated with the Coulomb 
charging energy due to the excess number of Cooper pairs within coarse-grained regions with
linear dimensions on the order of the superconducting coherence length. As we are primarily 
interested in the low temperature helicity moduli of this anisotropic, but three-dimensional, 
model we apply the variational self-consistent phonon approximation of Wood and Stroud [8]. 
The inhomogeneity of Josephson couplings and layer spacings perpendicular to the layers makes the 
calculation of the $z$-axis helicity modulus a nontrivial extension of the work by 
Roddick and Stroud [2]. Moreover we specifically address the analytic form (power 
law behavior) of the temperature dependence in the in-plane and out-of-plane 
superfluid densities by calculating the relevant weighted phase-phonon densities of states. 

The rest of the paper is organized as follows. In section 2 we discuss our model and 
derive the expressions for the helicity moduli within the self-consistent phonon approximation.
Section 3 contains our numerical results for the
helicity moduli, an analysis of their low temperature behavior including the relevant 
analytical results for weighted phase-phonon densities of states 
and our conclusions.
\section{The model and helicity moduli}

We consider an anisotropic quantum 3D XY-model for a system with bilayer structure
described by the Hamiltonian 
\begin{eqnarray}
{\hat H} & = & {\hat T}+{\hat V}_{ab}+{\hat V}_{c}\>,\\
{\hat T} & = & \frac{U}{2}\sum_{l,s,i}\left[-2i\frac{d}{d\phi_{i,l,s}}\right]^{2}\>,\\
{\hat V}_{ab} & = & \sum_{l,s,\langle i,j\rangle}\left[J_{1}(1-
\cos(\phi_{i,l,s}-\phi_{j,l,s}))\right]\>,\\
{\hat V}_{c} & = & \sum_{l,i}\left[J_{\perp}(1-
\cos(\phi_{i,l,2}-\phi_{i,l,1})) + 
J_{\perp}'(1-\cos(\phi_{i,l+1,1}-\phi_{i,l,2}))\right]\>.
\end{eqnarray}
Here, the sum over $l$ runs over a stack of bilayers, the sum over $s=1,2$ runs over two 
layers in a given bilayer, $\langle i,j\rangle$ denotes the nearest neighbors within a single 
layer, the sum over $i$ runs over the sites in a given layer and $\phi_{i,l,s}$ 
is the phase of the order parameter on site $i$ of the layer $s$ in the bilayer $l$.
${\hat T}$ is the charging energy associated with excess number of Cooper pairs on site $(i,l,s)$ [8], 
while ${\hat V}_{ab}$ and ${\hat V}_{c}$ describe the Josephson coupling energy within the layers and 
between the layers, respectively. The model considered here differs from the one examined
by Roddick and Stroud [2] in that the Josephson coupling constant between 
two layers within a given bilayer $J_{\perp}$ and their spacing $c_{b}$ are different 
from the Josephson coupling constant between layers in two adjacent bilayers $J_{\perp}'$ 
and the bilayer spacing $c'$. As a result, the calculation of the 
helicity modulus $\gamma_{zz}$ along direction perpendicular to the bilayers
is a nontrivial generalization of the work by Roddick and Stroud [2]. 

In this work we apply the self-consistent phonon approximation [8] to the low temperature   
thermodynamics implied by the Hamiltonian ${\hat H}$. One would expect that for a three-dimensional 
model this approach provides a reasonable description of the low temperature thermodynamics,  
at least for not too large values of $U/\max\{J_{1},J_{\perp},J_{\perp}'\}$. However, one cannot 
{\em a priori} rule out a possibility that quantum Monte Carlo treatment reveals, in 
particular in the limit of high anisotropy, a qualitatively different behavior 
of the type found by Jacobs {\em et al.} [9] in {\em purely two-dimensional} case.  
We leave quantum Monte Carlo treatment of our model for the future investigations.

In variational self-consistent phonon approach the trial free energy has the form 
\begin{equation}
F_{t}=F_{h}-\langle {\hat H}-{\hat H}_{h}\rangle_{h}\>, 
\end{equation}
where
\begin{eqnarray}
{\hat H}_{h} & = & {\hat T}+{\hat V}_{ab}^{(h)}+{\hat V}_{c}^{(h)}\>,\\
{\hat V}_{ab}^{(h)} & = & \sum_{l,s,\langle i,j\rangle}\frac{K_{1}(T)}{2}(\phi_{i,l,s}-\phi_{j,l,s})^{2}\>,\\
{\hat V}_{c}^{(h)} & = &\sum_{l,i}\left[\frac{K_{\perp}(T)}{2}(\phi_{i,l,2}-\phi_{i,l,1})^{2}+
\frac{K_{\perp}'(T)}{2}(\phi_{i,l+1,1}-\phi_{i,l,2})^{2}\right]\>,
\end{eqnarray}
$\langle\cdots\rangle_{h}$ denotes the statistical average with respect to the 
quasi-harmonic Hamiltonian ${\hat H}_{h}$ with temperature dependent ``spring constants'', 
and $F_{h}$ is the Helmholtz free energy defined by the Hamiltonian ${\hat H}_{h}$. The 
values of $K_{1}(T)$, $K_{\perp}(T)$ and $K_{\perp}'(T)$ are determined from self-consistency 
equations which result from minimizing the trial free energy $F_{t}$. 

Transforming to the normal modes and using $\langle\cos(\phi-\phi')\rangle_{h}={Re}\langle\exp(i
(\phi-\phi'))\rangle_{h}$ together with Mermin's theorem [10] we obtain
\begin{eqnarray}
\fl F_{t}=k_{B}T\sum_{s=1,2}\sum_{{\bf k}}\ln\left(2\sinh\frac{\hbar\omega_{s}({\bf k})}{2k_{B}T}\right)
+4N\left[J_{1}(1-e^{-\frac{1}{2}D_{1}})-\frac{1}{2}K_{1}D_{1}\right] \nonumber \\
+N\left\{\left[J_{\perp}(1-
e^{-\frac{1}{2}D_{\perp}})-\frac{1}{2}K_{\perp}D_{\perp}\right]+\left[J_{\perp}'(1-
e^{-\frac{1}{2}D_{\perp}'})-\frac{1}{2}K_{\perp}'D_{\perp}'\right]\right\}
\end{eqnarray}
and 
\begin{eqnarray}
K_{1}(T) & = & J_{1}e^{-\frac{1}{2}D_{1}(T)}\>,\\
K_{\perp}(T) & = & J_{\perp}e^{-\frac{1}{2}D_{\perp}(T)}\>,\\
K_{\perp}'(T) & = & J_{\perp}'e^{-\frac{1}{2}D_{\perp}'(T)}\>,
\end{eqnarray}
where
\begin{eqnarray}
\fl D_{1}(T) = \frac{1}{N}\sum_{s=1,2}\sum_{{\bf k}}\left(\sin^{2}(k_{x}a/2)
+\sin^{2}(k_{y}a/2)\right)
\frac{\hbar}{2M\omega_{s}({\bf k})}\coth\frac{\hbar\omega_{s}({\bf k})}{2k_{B}T}\>,\\
\fl D_{\perp}(T) =  \frac{1}{N}\sum_{s=1,2}\sum_{{\bf k}}\frac{\hbar}{2M\omega_{s}({\bf k})} \nonumber \\
\times\left[1+(-1)^{s}\frac{K_{\perp}+K_{\perp}'\cos(k_{z}c)}{\sqrt{(K_{\perp}+K_{\perp}')^{2}-
4K_{\perp}K_{\perp}'\sin^{2}(k_{z}c/2)}}\coth\frac{\hbar\omega_{s}({\bf k})}{2k_{B}T}\right]\>,\\
\fl D_{\perp}'(T) = \frac{1}{N}\sum_{s=1,2}\sum_{{\bf k}}\frac{\hbar}{2M\omega_{s}({\bf k})} \nonumber \\
\times\left[1+(-1)^{s}\frac{K_{\perp}\cos(k_{z}c)+K_{\perp}'}{\sqrt{(K_{\perp}+K_{\perp}')^{2}-
4K_{\perp}K_{\perp}'\sin^{2}(k_{z}c/2)}}\coth\frac{\hbar\omega_{s}({\bf k})}{2k_{B}T}\right]\>,\\
\fl \omega_{1,2} = \left\{\frac{2}{M}\left[2K_{1}\left(\sin^{2}(k_{x}a/2)
+\sin^{2}(k_{y}a/2)\right)+\frac{K_{\perp}+K_{\perp}'}{2}\right.\right. \nonumber \\
 \mp \left.\left.\frac{1}{2}\sqrt{(K_{\perp}+K_{\perp}')^{2}-4K_{\perp}K_{\perp}'\sin^{2}(k_{z}c/2)}\right]\right\}^{1/2}\>.
\end{eqnarray}
The sums over ${\bf k}$ run over the first Brillouin zone, $N$ is the number of unit cells and 
$M = \hbar^{2}/(4U)$. Equations (9-16) can be solved iterativly at fixed temperature for a given 
set of the bare coupling constants $J_{1}$, $J_{\perp}$, $J_{\perp}'$ and for a given 
Coulomb parameter $U$. In (16) index $1$ denotes the acoustic phase phonon dispersion and 
index $2$ denotes the optic one.

When a uniform vector potential ${\bf A}$ is applied its effect on the
Hamiltonian is to shift the phase difference between points ${\bf r}_{1}$ and 
${\bf r}_{2}$ by $A_{1,2}=2\pi {\bf A}\cdot({\bf r}_{2}-{\bf r}_{1})/\Phi_{0}$, 
where $\Phi_{0}=hc/2e$ is the flux quantum. For example, if ${\bf A}$ is 
perpendicular to the bilayers $\phi_{i,l,2}-\phi_{i,l,1}$ and 
$\phi_{i,l+1,1}-\phi_{i,l,2}$ in Eqs. (4) and (8) are replaced by 
$\phi_{i,l,2}-\phi_{i,l,1}+(2\pi/\Phi_{0})c_{b}A$ and 
$\phi_{i,l+1,1}-\phi_{i,l,2}+(2\pi/\Phi_{0})c'A$, respectively. As a result, for ${\bf A}$ 
perpendicular to the bilayers the quasi-harmonic Hamiltonian, (6-8), takes the form 
\begin{eqnarray}
{\hat H}_{h}(A) & = & {\hat H}_{h}(0)+{\hat H}'(A) \\
{\hat H}'(A) & = & \frac{2\pi}{\Phi_{0}}AN(K_{\perp}c_{b}-K_{\perp}'c')(\phi_{{\bf k}=0,2}- 
\phi_{{\bf k}=0,1})
\end{eqnarray}
where a constant term, proportional to $A^{2}$, has been dropped and $\phi_{{\bf k},s}$ 
is the Fourier transform of the phase variable $\phi_{{\bf l},s}$ defined on Bravais lattice 
sites ${\bf l}$ (s=1,2). 

The computation of the helicity modulus along direction perpendicular to the bilayers 
requires even more care than in the classical case [1] because $\partial 
{\hat H}_{h}(A)/\partial A$ does not commute with ${\hat H}_{h}(A)$ (see Eqs.~(17-18)) and 
therefore $\partial\exp(-\beta{\hat H}_{h}(A))/\partial A\neq -\beta(\partial{\hat H}_{h}(A)/
\partial A)\exp(-\beta{\hat H}_{h}(A))$. Hence the helicity modulus $d\langle {\hat J}(A)
\rangle_{h,A}/dA|_{A=0}$, where ${\hat J}(A)$ is the current operator through a particular 
bond perpendicular to the layers and $\langle\cdots\rangle_{h,A}$ is the usual quantum 
statistical average given by ${\hat H}_{h}(A)$, {\em is not} given by the standard 
fluctuation-type formula $\langle d{\hat J}(A)/dA|_{A=0}\rangle_{h,A=0}-(1/k_{B}T)
\langle{\hat J}(0)\partial{\hat H}_{h}(A)/\partial A|_{A=0}\rangle_{h,A=0}+(1/k_{B}T)
\langle{\hat J}(0)\rangle_{h,A=0}\langle\partial{\hat H}_{h}(A)/\partial A|_{A=0}\rangle_{h,A=0}$.

Instead, we compute the average current $\langle{\hat J}(A)\rangle_{h,A\neq0}$ to the  
lowest order in $A$ using the perturbation theory and evaluate the helicity modulus from 
$d\langle{\hat J}(A)\rangle_{h,A\neq0}/dA|_{A=0}$. Focusing on a bond between two layers 
in a bilayer (strong bond) we have [1]
\begin{equation}
{\hat J}_{s}(A)=J_{\perp}\sin(\phi_{i,l,2}-\phi_{i,l,1}+(2\pi/\Phi_{0})c_{b}A).
\end{equation}
Writing $\exp(-\beta{\hat H}_{h}(A))=\exp(-\beta{\hat H}_{h}(0){\hat U}(\beta,A)$ we get the 
equation of motion $\partial {\hat U}(\beta,A)/\partial \beta= {\hat H}'(\beta,A){\hat U}(\beta,A)$, ${\hat H}'(\beta,A)=\exp(\beta{\hat H}_{h}(0)){\hat H}'(A)\exp(-\beta{\hat H}_{h}(0))$, 
and the boundary condition ${\hat U}(0,A)=1$. Thus, to the lowest order in $A$ we have 
\[
{\hat U}(\beta,A)=1-\int_{0}^{\beta}d\tau{\hat H}'(\tau,A){\hat U}(\tau,A)\approx 
1-\int_{0}^{\beta}d\tau{\hat H}'(\tau,A)\>.
\]

Using this result, $\langle\sin(\phi-\phi')\rangle_{h,A\neq0}=(1/i){Im}\langle\exp(i
(\phi-\phi'))\rangle_{h,A\neq0}$ and the operator identities 
\begin{eqnarray}
{\hat b}e^{c({\hat b}+{\hat b}^{\dagger})} & = & \frac{1}{c}\frac{\partial}{\partial\alpha}
e^{c(\alpha{\hat b}+{\hat b}^{\dagger})}|_{\alpha=1}+\frac{c}{2}
e^{c({\hat b}+{\hat b}^{\dagger})} \\
{\hat b}^{\dagger}e^{c({\hat b}+{\hat b}^{\dagger})} & = & \frac{1}{c}\frac{\partial}{\partial\alpha}
e^{c({\hat b}+\alpha{\hat b}^{\dagger})}|_{\alpha=1}-\frac{c}{2}
e^{c({\hat b}+{\hat b}^{\dagger})} 
\end{eqnarray}
valid for any two operators ${\hat b}$ and ${\hat b}^{\dagger}$ such that 
$[{\hat b},{\hat b}^{\dagger}]=1$, we find to the lowest order in $A$
\begin{equation}
\langle{\hat J}_{s}(A)\rangle_{h,A\neq0}=A\frac{2\pi}{\Phi_{0}}\frac{K_{\perp}(T)K_{\perp}'(T)}
{K_{\perp}(T)+K_{\perp}'(T)}(c_{b}+c')\>.
\end{equation}
Thus the $z$-axis helicity modulus of our model in quasi-harmonic approximation is given 
by
\begin{equation}
\gamma_{zz}(T)=\frac{2\pi}{\Phi_{0}}\frac{K_{\perp}(T)K_{\perp}'(T)}
{K_{\perp}(T)+K_{\perp}'(T)}(c_{b}+c')\>.
\end{equation}
Needless to say, we get the same result for $\gamma_{zz}(T)$ by computing the average 
current along the weak bond, i.e. the one between the bilayers, as expected from Kirchhoff's 
first law (note the symmetry of the expression (23) under the exchange of
$(K_{\perp},c_{b})$ and $(K_{\perp}',c')$). 

Another way to write our central result Eq.~(23), which is intuitively appealing, 
would be 
\begin{equation}
\gamma_{zz}(T)  =  \frac{2\pi}{\Phi_{0}}K_{\perp,eff}(T)(c_{b}+c')\>, \\
\end{equation}
where
\begin{equation}
\frac{1}{K_{\perp,eff}}  =  \frac{1}{K_{\perp}}+\frac{1}{K_{\perp}'}\>.
\end{equation}
Equation (25) gives the effective spring constant for two linear springs with 
constants $K_{\perp}$ and $K_{\perp}'$ which are connected in series. 
The lattice parameter in direction perpendicular to the layers is $c_{b}+c'$.
The form given by Eqs.~(24-25) is analogous to what one finds  
using quasi-harmonic approximation for a lattice
without basis [2], e.g.~for the in-plane helicity modulus of our model
\begin{equation}
\gamma_{xx}(T)  =  \frac{2\pi}{\Phi_{0}}K_{1}(T)a\>.
\end{equation}
In that case the helicity modulus is essentially the stiffness with 
respect to the phase twist on neighboring sites. We note that within the 
quasi-harmonic approximation $\gamma_{zz}(T)$ does not depend on the relative 
size of the bond lengths $c_{b}$ and $c'$ (see Eqs.~(10-16), (23)), and the 
$z$-axis lattice parameter $c=c_{b}+c'$ enters only implicitly as $k_{z}c$ 
in the Brillouin zone sums.

In conclusion of this section we note that if one takes first the classical 
limit $\hbar\rightarrow0$ and then the zero temperature limit $T\rightarrow0$ 
in Eqs.~(10-16), (23) one recovers the classical results obtained previously 
for the bilayer model using the low temperature spin-wave expansion [1].

\section{Numerical results and conclusions}

We have computed the helicity moduli $\gamma_{zz}(T)$ and $\gamma_{xx}(T)$ 
for different choices of the bare coupling constants $J_{1}$, $J_{\perp}$, $J_{\perp}'$ and since
they give qualitatively very similar results we present here only the data for a 
single set of parameters $J_{1}$ = 1, $J_{\perp}$ = 0.1, and  $J_{\perp}'$ = 0.001. 
The value of Josephson coupling between the layers of a bilayer is taken to be one order of 
magnitude smaller than the in-plane Josephson coupling and the value of Josephson coupling 
between bilayers $J_{\perp}'$ is two orders of magnitude smaller than $J_{\perp}$.
We have considered five values of the Coulomb repulsion parameter $U$ = 10$^{-6}$, 10$^{-3}$, 
10$^{-2}$, 10$^{-1}$ and 1 (in units of the $ab$-plane coupling $J_{1}$).  
Since the helicity modulus is proportional to inverse square of the magnetic field  
penetration depth $\lambda(T)$ [2], we present our results for the helicity moduli in Fig.~1 
as $\lambda^{2}(0)/\lambda^{2}(T)$ as a function of $T$. We did not attempt to determine 
the transition temperature $T_{c}$ for various values of $U$ as it is unlikely that the 
self-consistent phonon approximation treats correctly vortex-antivortex pairs close to $T_{c}$ 
and we are primarily interested in the low temperature dependence of the helicity moduli. 

\begin{figure}
\begin{center}
\epsfbox{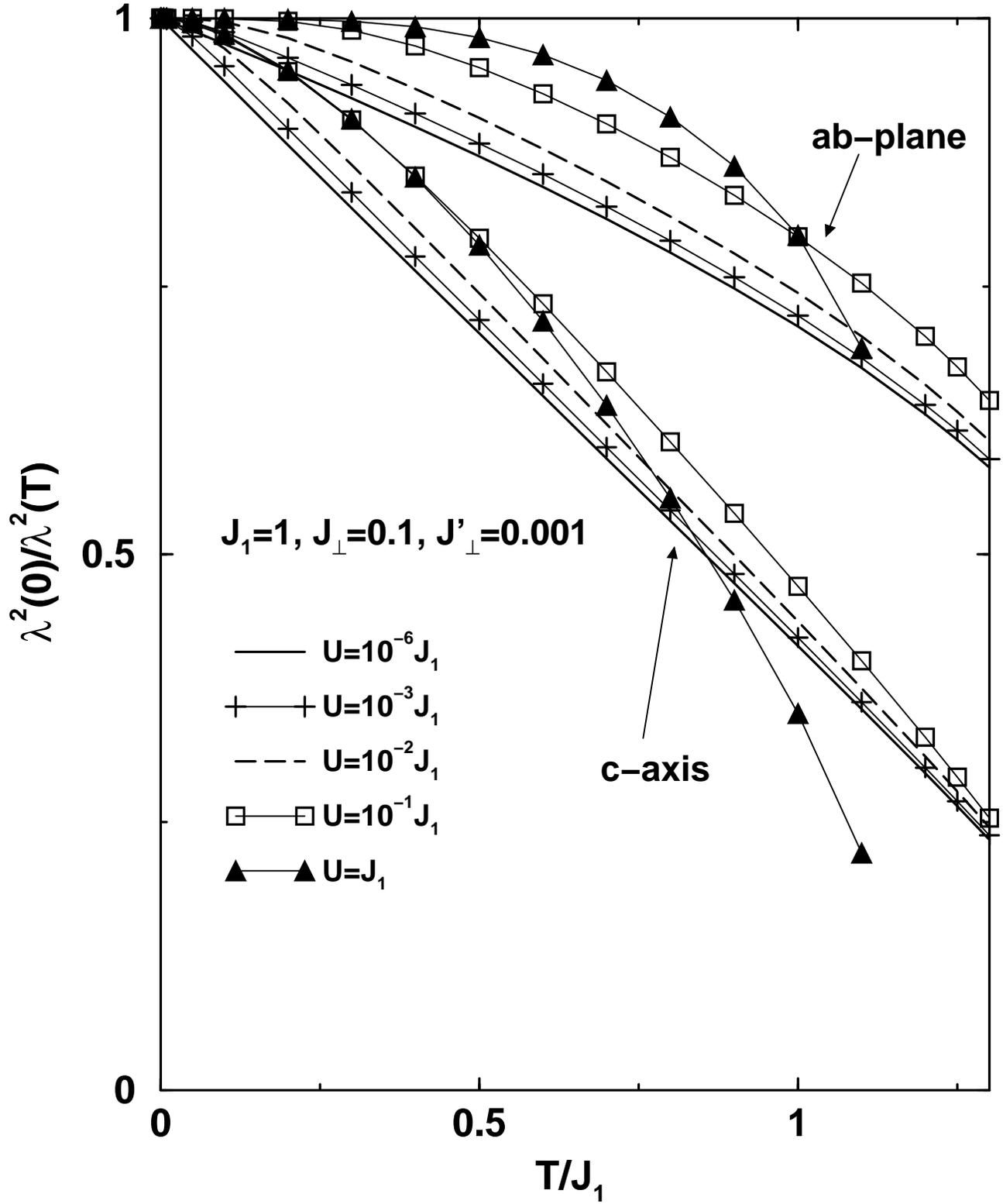}
\end{center}
\caption{\label{label}The temperature dependence of the helicity moduli for different values 
of the Coulomb repulsion parameter $U$. The top group of five curves are the $ab$-plane helicity 
moduli for different values of $U$, and the bottom group of curves are the corresponding 
$c$-axis helicity moduli.} 
\end{figure}

We point out that in the zero temperature limit 
$\gamma_{zz}(T)$ and $\gamma_{xx}(T)$ are {\em not linear} in $T$ 
for any finite $U$. It is easy to show that in the limit $T\rightarrow0$ both helicity 
moduli vary as $T^{4}$. Indeed, the expressions for $D_{1}(T)$, $D_{\perp}(T)$ and 
$D_{\perp}'(T)$, which determine the temperature dependences of $\gamma_{xx}(T)$ and 
$\gamma_{zz}(T)$, can be conveniently written in terms of various weighted phase-phonon 
densities of states
\begin{eqnarray}
\fl D_{1}(T)  =  \int_{0}^{+\infty}d\omega[F_{1}^{(1)}(\omega)+F_{1}^{(2)}(\omega)]\frac{\hbar^2}{2M\omega}\coth\frac{\omega}{2k_{B}T}\>, \\
\fl F_{1}^{(s)}(\omega)  =  \frac{v}{(2\pi)^{3}}\int_{FBZ}d^{3}{\bf k}\left(\sin^{2}(k_{x}a/2)
+\sin^{2}(k_{y}a/2)\right)\delta^{(3)}(\omega-\hbar\omega_{s}({\bf k}))\>,\>\>s=1,2\>, \\
\fl D_{\perp}(T)  =  \int_{0}^{+\infty}d\omega[F_{\perp}^{(1)}(\omega)+F_{\perp}^{(2)}(\omega)]
\frac{\hbar^{2}}{2M\omega}\coth\frac{\omega}{2k_{B}T}\>, \\
\fl F_{\perp}^{(s)}(\omega)  =  \frac{v}{(2\pi)^{3}}\int_{FBZ}d^{3}{\bf k}
\left[1+(-1)^{s}\frac{K_{\perp}+K_{\perp}'\cos(k_{z}c)}{\sqrt{(K_{\perp}+K_{\perp}')^{2}-
4K_{\perp}K_{\perp}'\sin^{2}(k_{z}c/2)}}\right] \nonumber \\
\times\delta^{(3)}(\omega-\hbar\omega_{s}({\bf k}))\>,\>\>s=1,2\>,\\
\fl D_{\perp}'(T)  =  \int_{0}^{+\infty}d\omega[F_{\perp}^{'(1)}(\omega)+F_{\perp}^{'(2)}(\omega)]
\frac{\hbar^2}{2M\omega}\coth\frac{\omega}{2k_{B}T}\>, \\
\fl F_{\perp}^{'(s)}(\omega)  =  \frac{v}{(2\pi)^{3}}\int_{FBZ}d^{3}{\bf k}
\left[1+(-1)^{s}\frac{K_{\perp}\cos(k_{z}c)+K_{\perp}'}{\sqrt{(K_{\perp}+K_{\perp}')^{2}-
4K_{\perp}K_{\perp}'\sin^{2}(k_{z}c/2)}}\right] \nonumber \\
\times\delta^{(3)}(\omega-\hbar\omega_{s}({\bf k}))\>,\>\>s=1,2\>,
\end{eqnarray}
where $v=a^2c$ is the volume of the unit cell.

We have computed $F_{1}^{(s)}$, $F_{\perp}^{(s)}$, $F_{\perp}'^{(s)}$, $s$ =1,2, numerically using 
the tetrahedron method [11]. 
At low temperatures it is the frequency dependence of the acoustic phase phonon 
weighted densities of states $F_{1}^{(1)}$, $F_{\perp}^{(1)}$ and $F_{\perp}'^{(1)}$ 
that determines the temperature dependence of $D_{1}(T)$, $D_{\perp}(T)$, $D_{\perp}'(T)$, 
and thereby of $K_{1}(T)$, $K_{\perp}(T)$, $K_{\perp}'(T)$ (Eqs.~(10-12)), which in 
turn determine the temperature dependences of the helicity moduli (Eqs.~(23) and (26)). 
In the limit $\omega\rightarrow0$ we find that because of the weighing factors 
$F_{1}^{(1)}$, $F_{\perp}^{(1)}$ and $F_{\perp}'^{(1)}$ are all proportional to $\omega^{4}$ 
(the coefficients are $a_{1}=\sqrt{2(K_{\perp}+K_{\perp}')/(UK_{\perp}K_{\perp}')}/$2$^{7}\pi^{2}$3
$(UK_{1})^{2}$, $a_{\perp}=a_{1}\left(2(K_{1}/(K_{\perp}+K_{\perp}')(K_{\perp}'/K_{\perp}))\right)$, and 
$a_{\perp}'=a_{\perp}(K_{\perp}/K_{\perp}')^{2}$, respectively). Then, it is easy to see, by rescaling the 
integration variables in Eqs.~(27), (29), (31), that $D_{1}(T)-D_{1}(0)\propto T^{4}$, etc..  
After expanding, one finds that both helicity moduli are decreasing with $T$ as $T^{4}$ in the 
zero temperature limit. We note that Xiang and Wheatley [12] found $T^{5}$-dependence of the $c$-axis superfluid density, which 
is proportional to $\gamma_{zz}(T)$, due to 
{\em entirely different} physical mechanism than the one considered in the present work. 

The results shown in Fig.~1 are qualitatively different from what is found experimentally on 
single crystals of bilayer compound of optimally doped YBCO [7] (see Fig.~2 in [7]). 
While experimentally the $ab$-plane helicity moduli are linear in $T$ at low temperatures we 
obtain $T^{4}$-dependence in the limit $T\rightarrow0$ for any 
finite U. Roddick and Stroud [2] found (for a one-layer model of YBCO) that including 
dissipation through coupling to an Ohmic heat bath restores the linearity of helicity 
moduli (presumably in {\em all} directions). We did not consider such coupling in this work. 
Physically the dissipation results from quasiparticle degrees of freedom [13], and we have 
discovered recently [14] that the effect of nodal quasiparticles on phase fluctuations in a 
d-wave superconductor might render the XY-model type of description of such system 
meaningles in the limit $T\rightarrow0$. Nevertheless, the biggest discrepancy between the 
experiments [7] and our results in Fig.~1 is that the calculated $\gamma_{zz}(T)/\gamma_{zz}(0)$ 
decreases much faster with temperature than the calculated $\gamma_{xx}(T)/\gamma_{xx}(0)$. This 
feature was also present in our Monte Carlo study of the classical version ($U$ = 0) of the 
same model [1] and in either case it is a direct consequence of the weaker Josephson couplings 
in direction perpendicular to the layers compared to the in-plane Josephson coupling. Dissipation, 
treated as in [2], will not affect that result. 

In conclusion, neither a classical nor a quantum-mechanical XY-model (at least in the self-cosistent 
phonon approximation) can consistently account for both $ab$-plane and $c$-axis electrodynamics of 
YBCO observed in experiments [7].

\ack
This work has been supported in part by the Natural Sciences and Engineering 
Research Council of Canada.
\end{document}